\documentclass[10pt]{article}

\usepackage{graphicx}
\usepackage{amsmath}
\usepackage{amsfonts}
\usepackage{color}
\usepackage{times}
\usepackage{overpic}        

\def\R{\mathbb{R}}

\def\N{\mathbb{N}}
\def\Z{\mathbb{Z}}

\def\cI{\mathcal{I}}

\def\cO{\mathcal{O}}

\def\cS{\mathcal{S}}

\def\txtc{{\textnormal{c}}}
\def\txtd{{\textnormal{d}}}
\def\txte{{\textnormal{e}}}
\def\txti{{\textnormal{i}}}

\def\txtD{{\textnormal{D}}}

\newcommand{\be}{\begin{equation}}
\newcommand{\ee}{\end{equation}}
\newcommand{\benn}{\begin{equation*}}
\newcommand{\eenn}{\end{equation*}}
\newcommand{\bea}{\begin{eqnarray}}
\newcommand{\eea}{\end{eqnarray}}
\newcommand{\beann}{\begin{eqnarray*}}
\newcommand{\eeann}{\end{eqnarray*}}

\newcommand{\xt}{z}
\newcommand{\xs}{x^*}

\newcommand{\hl}[1]{#1}

\title{A Universal Route to Explosive Phenomena}

\author
{
Christian Kuehn$^{1,2}$ and Christian Bick$^{3,4,5,6\ast}$\\
\\
\normalsize{$^1$Faculty of Mathematics, Technical University of Munich, Garching, Germany}\\
\normalsize{$^2$Complexity Science Hub Vienna, Vienna, Austria}\\
\normalsize{$^3$Department of Mathematics, Vrije Universiteit Amsterdam, Amsterdam, the Netherlands}\\
\normalsize{$^4$Institute for Advanced Study, Technical University of Munich, Garching, Germany}\\
\normalsize{$^5$Mathematical Institute, University of Oxford, Oxford, UK}\\
\normalsize{$^6$Department of Mathematics, University of Exeter, Exeter, UK}\\
\normalsize{$^\ast$To whom correspondence should be addressed; E-mail: c.bick@vu.nl.}
}

\date{}

\begin{document} 




\maketitle

\begin{abstract}
Critical transitions are observed in many complex systems. This includes the onset of synchronization in a network of coupled oscillators or the emergence an epidemic state within a population. ``Explosive'' first-order transitions have caught particular attention in a variety of systems when classical models are generalized by incorporating additional effects. Here we give a mathematical argument that the emergence of such first-order transitions is not surprising but rather a universally expected effect: Varying a classical model along a generic two-parameter family must lead to a change of the criticality. To illustrate our framework, we give three explicit examples of the effect in distinct physical systems: a model of adaptive epidemic dynamics, for a generalization of the Kuramoto model, and for a percolation transition.   
\end{abstract}

\section*{Introduction}

Many complex nonlinear systems---ranging from epidemic spreading, synchronization of coupled oscillators, to percolation on a network---undergo critical order-disorder transitions as a system parameter is varied.
As in classical statistical mechanics~\cite{Haken1977}, these transitions can be continuous (second-order) or discontinuous (first-order) at the transition point. Discontinuous first-order transitions have attracted particular attention~\cite{Boccaletti2016,DSouza2019} as they can lead to an ``explosive'' change of system properties. For a wide variety of complex systems it has been observed that a variation of the model via additional features leads to the change from a continuous second-order to a discontinuous first-order critical transition. 
For example, the classical Kuramoto model shows a continuous synchronization transition. However, varying the distribution of intrinsic frequencies~\cite{Pazo2005,OmelChenko2012a}, generalizing the network to simplicial or higher-order coupling~\cite{Skardal2019b,Millan2019a}, \hl{or adding a dynamical rule that suppresses synchronization~\cite{Zhang2014}} all allow for discontinuous synchronization transitions. 
Similarly, adding adaptation~\cite{GrossDLimaBlasius} or higher-order coupling structures~\cite{Iacopinietal} to models of epidemic spreading can induce a discontinuous transition to the epidemic state.
These and other examples follow the same paradigm: First, an additional effect is added to a classical model. Second, variation of a parameter associated with the new effect turns a previously second-order
transition into a first-order
transition.%

Here we give a mathematical argument that a transition from a continuous to a discontinuous critical transition is not surprising
in nonlinear dynamical systems but a generically/universally expected effect if additional parameters are varied. Specifically, we show that \emph{any} typical model variation along a two-parameter family of any classical model with a second-order transition must lead to a change of the criticality to first-order.
First, this result shows that a change of the criticality of transition points in different complex systems has a common dynamical origin; we illustrate this in explicit examples involving adaptive epidemic dynamics, synchronization in the Kuramoto model with non-additive higher-order interactions, and a model from percolation theory.
Second, this insight can be useful to identify system perturbations to induce or prevent the emergence of abrupt critical transitions in a variety of physical systems.
\hl{Third, it highlights that means to delay the onset of a critical transition---such as adding a suppressive rule---can change the nature of a critical transition from being continuous to being discontinuous.}

\section*{Results}

\subsection*{A universal mechanism that modulates transitions}

\subsubsection*{Normal forms for critical transitions}
\hl{Here we take a dynamical approach at critical transitions where the macroscopic dynamics change qualitatively; we relate the Landau's classical approach to phase transitions below.}
Consider a high-dimensional dynamical system close to a critical transition point. Mean-field approximations are a commonly used tool to simplify such a system to a low-dimensional description in terms of mean-field variables. Such approximations can be obtained through moment closure and other approaches; see~\cite{KuehnMC,Bick2018c} and the examples below. Consequently, we assume that the  mean-field or continuum limit dynamics of the physical system in question near the transition point are given by an ordinary differential equation (ODE)
\be
\label{eq:ODE}
\dot x :=\frac{\txtd x}{\txtd t}=F(x,y),
\ee
where $x=x(t)\in\R^n$ are the mean-field variables and $y\in\R^m$ are model parameters.
While the state undergoing a transition in the full system can be quite general---that is, it does not necessarily have to be a stationary solution \hl{of the microscopic dynamics}---we assume it corresponds to an equilibrium point for the macroscopic equations~\eqref{eq:ODE}: Hence, suppose that $\xs=\xs(y)$ is the corresponding smooth family of equilibrium points parametrized by~$y$ that exist for all parameters.
Using a translation, we may shift the equilibria and assume that $\xs= 0:=(0,0,\ldots,0)^\top\in\R^n$ is the equilibrium point, i.e., $F(0,y)=0$ for all $y\in\R^m$. 

Now suppose that the transition point corresponds to a bifurcation point~\cite{GH} upon parameter variation. Generically, we can assume that a single eigenvalue of the Jacobian
$\txtD_x F(0,y)\in\R^{n\times n}$
crosses the imaginary axis.
First, by a translation in parameter space we may assume that the main bifurcation parameter is $p=y_1$.
Second, using center manifolds~\cite{GH} or Lyapunov--Schmidt reduction~\cite{Kielhoefer}, we have that the dynamics of the full system~\eqref{eq:ODE} are locally given by the one-dimensional dynamics on the center manifold,
\be
\label{eq:ODE1}
\dot x=f(x,p),\quad x\in\R,~p\in\R,
\ee
with a family of equilibria~$\xs(p)=0$, that is,
\be
\label{eq:FP1}
f(0,p)=0,\quad~p\in\R,
\ee
that bifurcate at $p=0$. Note that center manifold reductions do not necessarily require a finite-dimensional approximation~\eqref{eq:ODE} since it can also be directly applied to infinite-dimensional systems, such as partial differential equations or systems with delay; see, for example,~\cite{Vanderbauwhede1992a,Faye2018}.

Now, it is well-known in bifurcation theory, that the two typical bifurcation points encountered in applications are the transcritical bifurcation with local normal form
\be
\label{eq:tc}
\dot x = p x + a x^2,
\ee
where $a=\pm 1$ determines whether the bifurcation/transition upon varying~$p$ is second-order 
($x\geq 0$, $a=-1$) or first-order ($x\geq 0$, $a=+1$). Similarly, if there is an equivariance 
given by a $\Z_2$-reflection symmetry in the model via $f(x,p)=-f(-x,p)$, \hl{then there cannot be any terms of even power in~$x$ in the Taylor expansion} and the generic transition is a pitchfork bifurcation
\be
\label{eq:pitch}
\dot x = p x + a x^3.
\ee
The pitchfork is second-order if it is supercritical and $a=-1$, while it is first-order 
if it is subcritical and $a=+1$.

\newcommand{\Xs}{X^*}

\hl{This dynamical perspective directly relates to Landau's classical theory of phase transitions; cf.~\cite{Landau2013}. In brief, this approach relies on defining an order parameter~$X$ and constructing a functional~$G(X)$ whose minima~$\Xs$ are the values the free energy of the system takes. For the Ising model of interaction spins, the order parameter~$X$ is the average of all spins and the free energy functional is $G(X) = \rho X^2 + aX^4$, where~$\rho$ is a shifted temperature and $a\neq 0$ a parameter. By differentiation, we have that the minima~$\Xs$ that determine the free energy satisfy
\[\rho \Xs + 2a({\Xs})^3 = 0,\]
which corresponds to equilibria of the normal form of the pitchfork bifurcation~\eqref{eq:pitch}. Hence, Landau's phase transitions can analyzed by looking at the branching behavior of equilibria~\eqref{eq:FP1} in a corresponding dynamical model.
}

\subsubsection*{Change of criticality for a generic variation of an additional parameter}

\hl{We now consider the variation of an additional parameter that takes into account the additional effect for each model as indicated above.}
Take a generic single-eigenvalue crossing and with the phase space $\{x\geq 0\}$. As the model is varied, the persistence of a single-eigenvalue crossing is generic within one-parameter families of the vector field $f$. Hence, we may assume (without loss of generality) that the 
single eigenvalue crosses at $p=0$. Furthermore, if we vary the model at least one additional free parameter, say $q=y_2$, generically appears.

\hl{To understand how varying the additional parameter affects the equilibria, we now expand the vector field~\eqref{eq:ODE1} in~$x$ and~$p$ as well as~$q$.}
A Taylor expansion at the bifurcation point yields
\benn
f(x,p,q)=\sum_{r=0}^M\sum_{j+k+l=r} c_{jkl} x^j p^k q^l + \cO(M+1),
\eenn 
where $\cO(M+1)$ denotes terms of order $M+1$. \hl{The coefficients~$c_{jkl}$ are constrained by the conditions imposed by the equilibrium and the bifurcation scenario we consider.} 
First, the existence 
of a trivial branch of equilibria, $f(0,p,q)=0$, implies $c_{0kl}=0$ for all $k,l\in\N_0=\N\cup\{0\}$. 
Second, since a single eigenvalue crosses at $p=0$, we must have $\partial_xf(0,0,q)=0$, where 
$\partial_x$ denotes the partial derivative with respect to~$x$. Hence, we have $c_{10l}=0$ for all $l\in\N_0$. \hl{Third, since we assume a simple eigenvalue crosses transversally, we get $\partial_{xp}(0,0,q)\neq 0$ entailing $c_{110}\neq0$. In summary, we have}
\be
\label{eq:Taylor}
f(x,p,q)=c_{110} x p + c_{200} x^2 + \sum_{j+k+l=3} c_{jkl} x^j p^k q^l + \cO(4).   
\ee

\begin{figure}
	\centering
	\begin{overpic}[width=0.4\textwidth]{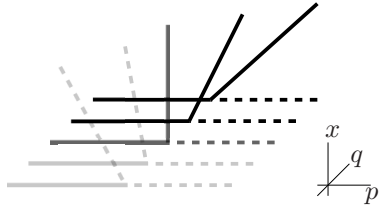}
	\put(98,0){$p$}
	\put(94,10){$q$}
	\put(87,17){$x$}
	\end{overpic}
	\caption{\label{fig:01}Sketch for the variation of a transcritical bifurcation
	for the phase space $\{x\geq 0\}$ and parameters $(p,q)$ with primary parameter $p$
	and second generic unfolding parameter $q$. Dashed lines indicate instability of 
	the equilibrium and solid lines indicate stability. The grey cases are first-order
	(subcritical) transitions, while the black diagrams are second-order 
	(supercritical) transitions.}
\end{figure} 

\hl{A generic model variation with at least one additional free parameter now leads to a vector field~$f$ that allows for a change in criticality. With the bifurcation conditions incorporated into~\eqref{eq:Taylor},} one may use bifurcation theory to unfold the singular point into a generic family. 
In particular, the next derivatives of the vector field at the bifurcation point should not vanish. Hence, for combinations with $j+k+l=3$ we must have $c_{102}=c_{0kl}=0$ from above and $c_{jkl}\neq 0$ if $j\geq 1$. The leading-order non-vanishing conditions are $\partial_{xxp}f(0)\neq 0$ and $\partial_{xxq}f(0)\neq 0$
\hl{and we note that~$c_{111} x pq$ is of higher order in comparison to~$c_{110} x p$ for the linear part in $x$ since $c_{110}\neq 0$.}
Truncating higher-order terms, this yields the lowest-order two-parameter unfolding normal form
\be
\label{eq:TCnf1}
f(x,p,q)=c_{110} x p + (c_{200} + c_{210}p+c_{201}q) x^2.
\ee
We now apply a scaling (or geometric desingularization, or renormalization) with a small
parameter $\varepsilon>0$ through the transformation
$(x,p,q)\mapsto (x\varepsilon^\alpha,p\varepsilon^\beta,q\varepsilon^\gamma)$.
For the transcritical normal form~\eqref{eq:TCnf1} we choose $\alpha=1$, $\beta=-1$, 
$\gamma=-2$ to obtain (upon a suitable time rescaling)
\be
\label{eq:TCnf2}
f(x,p,q)=c_{110} x p + (c_{200} \varepsilon^2 + c_{210}p \varepsilon +c_{201}q) x^2.   
\ee
Hence, one easily checks that there is a sign change of $\partial_{xx}f(0,p,q)$ 
upon varying $q$ in an interval $[-q_0,q_0]$ for some $q_0>0$ as long as $c_{201}\neq0$,
\hl{which we expect generically as it is the leading-order term involving the parameter $q$}. 
Even if $c_{201}=0$, we can expand to higher order in~$q$ 
and may thereby eventually change the sign of $\partial_{xx}f(0,p,q)$. So only certain
situations, e.g., the presence of symmetries or non-generic smooth functions could lead to the 
preservation of the sign for all $q\in\R$: 
\hl{A function without any dependence on the second parameter~$q$ may be the most extreme case of non-genericity, but symmetries can also force specific Taylor coefficients to vanish.}
Once the sign of $\partial_{xx}f(0,p,q)$ changes,  
this implies that generically the second parameter is able to change the transition from 
second order to first order or vice versa. Of course, from the viewpoint of the geometry of the 
bifurcation diagram, this is quite intuitive as shown in Fig.~\ref{fig:01} that a
second generic parameter may change criticality.

\begin{figure}
	\centering
	\begin{overpic}[width=0.4\textwidth]{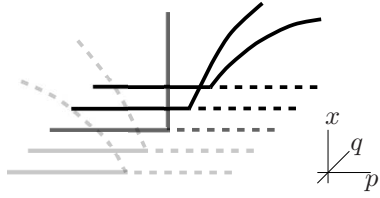}
	\put(98,0){$p$}
	\put(94,10){$q$}
	\put(87,17){$x$}
	\end{overpic}
	\caption{\label{fig:02}Sketch for the variation of a pitchfork bifurcation
	for the phase space $\{x\geq 0\}$ and parameters $(p,q)$ with primary parameter $p$
	and second generic unfolding parameter $q$. Dashed lines indicate instability of 
	the equilibrium and solid lines indicate stability. The grey cases are first-order
	(subcritical) transitions, while the black diagrams are second-order 
	(supercritical) transitions.}
\end{figure} 

The situation for the pitchfork works very similarly except that an additional
symmetry $f(x,p,q)=-f(-x,p,q)$ has to be respected. This further constrains 
the coefficients of the Taylor expansion. Note that if this symmetry is broken
then we are in the transcritical case if there is still a trivial branch for 
all values of the parameters. Hence, we now assume that the symmetry holds.
Taylor expansion as above gives for a bifurcation point with a single 
eigenvalue crossing   
\benn
f(x,p,q)=c_{110} x p + c_{300} x^3 + \sum_{j+k+l=4} c_{jkl} x^j p^k q^l + \cO(5).   
\eenn 
The same steps as above lead to leading-order to the two-parameter normal form
\be\label{eq:NFsym}
f(x,p,q)=c_{110} x p + (c_{300}+c_{310}p+c_{301}q) x^3.   
\ee 
Again, this shows that a second parameter can generically change a second-order to a 
first-order transition; cf.~Fig.~\ref{fig:02}.


\subsection*{Discontinuous critical transitions from a universal perspective}

We now give three explicit examples of complex nonlinear systems where a generalization leads to a change from a first-order to a second-order critical transition. While seemingly distinct and from a variety of contexts, we show that the transitions are related through the abstract framework above.

\subsubsection*{Transitions in adaptive epidemics}

\hl{Epidemic dynamics on complex networks has been a very active topic for several decades~\cite{Pastor-Satorrasetal}.
Classical Susceptible--Infected--Susceptible (SIS) models are microscopically modeled as a Markov chain on networks with nodes being in two states, either susceptible~$\cS$ or infected~$\cI$.
Infections take place at rate~$\rho$ along network links, recovery at rate~$r$ (which we set to $r=1$ without loss of generality here).
The effect of the underlying network is crucial to understand such contact processes~\cite{Pastor-SatorrasVespignani}, for example, to determine the fraction of infected individuals in the long run.}

\hl{However, changing social contacts interact affect the network and thus network adaptivity---dynamics \emph{of} the network that interact with the dynamics \emph{on} the network---affect epidemic dynamics. 
The paradigmatic and widely-used adaptive epidemic model by Gross et al.~\cite{GrossDLimaBlasius} considers SIS dynamics with additional adaptive re-wiring of an $\cS\cI$-link to an $\cS\cS$-link at rate~$q$.
While this rule may suppress infection (as $\cS\cI$-links are removed), it can also create highly connected clusters of susceptible nodes (as $\cS\cS$-link are added).} 
Direct numerical simulations show that the bifurcation at the epidemic threshold $\rho=\rho_\txtc$ is a second-order transition if $q=0$.
It becomes a first-order transition if~$q$ is increased sufficiently, i.e., the network 
becomes more strongly adaptive. Based upon our considerations above, it is natural to 
expect that allowing for general network topologies via re-wiring is a sufficiently generic 
breaking mechanism to allow the second-to-first order change via the parameter~$q$. In
fact, this is what is verified implicitly in~\cite{GrossDLimaBlasius} by using a moment-closure 
expansions~\cite{KuehnMC} of the network dynamics. 
The dynamics for large networks are described by the moment-closed ODEs
\begin{align*}
\dot I &= \rho\left(\frac{\mu}{2}- l_{\cI\cI}-l_{\cS\cS}\right)-I,\nonumber\\
\dot l_{\cI\cI} &= \rho\left(\frac{\mu}{2}- l_{\cI\cI}-l_{\cS\cS}\right)\left(\frac{\frac{\mu}{2}- l_{\cI\cI}-l_{\cS\cS}}{1-I}+1\right)-2 l_{\cI\cI},\\
\dot l_{\cS\cS} &= (1+q) \left(\frac{\mu}{2}- l_{\cI\cI}-l_{\cS\cS}\right)-\frac{2\rho(\frac{\mu}{2}- l_{\cI\cI}-l_{\cS\cS})l_{\cS\cS}}{1-I},\nonumber 
\end{align*}
where $I$ and $l_{\cI\cI}$, $l_{\cS\cS}$ are a normalized infected density and two similarly 
normalized link densities respectively~\cite{GrossDLimaBlasius}; note that conservation laws allow 
for the elimination of $S$ and $l_{\cS\cI}$. We fix $\mu$ arising 
from a connectivity assumption~\cite{GrossDLimaBlasius} of the network to $\mu=20$. This is a standard 
assumption~\cite{KuehnCT2}, as we only want to demonstrate the principal effect of adding re-wiring via $q$.
It can be checked, see~\cite{GrossDLimaBlasius,KuehnCT2}, that a first-order 
transition is possible upon varying~$q$.

We now formally show that the change of criticality is a special case of our more general results above. 
One checks that there always exists the invariant trivial branch of steady states $\{I=0,l_{\cI\cI}=0,
l_{\cS\cS}=\frac{\mu}{2}\}$. The epidemic threshold bifurcation point is given by
\benn
\rho_\txtc=\frac{1+q}{\mu}=\frac{1+q}{20}.
\eenn  
We now compute the normal form using a direct and general center manifold calculation (see~\cite{GH} for a general reference), which we outline here: First, we shift coordinates $I=X_1$, $l_{\cI\cI}=X_2$, 
$l_{\cS\cS}=X_3+10$, $\rho=p+\rho_c$, to obtain a vector field $\dot X=F(X,p,q)$. Then we transform 
the linear part $A=\txtD_{X}F(0,0,q)$ into Jordan canonical form
\benn
M^{-1}AM=\left(
\begin{array}{ccc}
 0 & 0 & 0 \\
 0 & -1 & 0 \\
 0 & 0 & \frac{1}{20} (-q-41) \\
\end{array}
\right)
\eenn 
for a transformation matrix $X=M\xt$ that can be calculated from the eigenvectors of $A$.
We augment the new ODEs $\dot\xt= M^{-1}F(M\xt,p,q)$ by $\dot p=0$ and $\dot q=0$ to calculate a 
the three-dimensional center manifold $\{(\xt_2,\xt_3)=h(\xt_1,p,q)\}$ as there 
are three zero eigenvalues. The manifold is parametrized over the center directions $(\xt_1,p,q)$. 
Using the invariance equation~\cite{GH} and a quadratic ansatz for $h$, one obtains after equating 
coefficients
\begin{align*}
\xt_2&=h_1(\xt_1,p,q)=-\frac{4}{35301}\xt_1^2,\\
\xt_3&=h_2(\xt_1,p,q)=\frac{3364}{206763}\xt_1^2+\frac{280}{1681}\xt_1p.
\end{align*}
Substituting this back into the equation for~$\dot\xt_1$ and writing $x:=\xt_1$ gives the flow 
on the center manifold to leading order as
\benn
\dot x=\frac{800}{q+41}xp + \frac{80 \left(\frac{2 (q+1)^2}{q+41}-\frac{1}{10} (q+1)\right)}{q+41} x^2+\dotsb.
\eenn
The coefficients of~$xp$ and~$x^2$ now show that the parameter~$q$ indeed yields a change in the criticality from a second-order to a first-order transition at $q=21/19$.
Hence, from this perspective we can clearly see that a change in criticality is not surprising: The re-wiring~$q$ appears in the reduced center manifold as a sufficiently generic second unfolding parameter as in the universal route described above.

\hl{Of course, there are several other possible variations of this theme for concrete models of contact process character that employ a secondary unfolding parameter to allow for a change criticality. Examples are epidemic models with additional node states~\cite{Janssen2004} or chemical reaction processes~\cite{Grassberger1982}; the change of criticality corresponds to a ``tricritical point''.
Yet, our theoretical results clearly show that one should be able to go far beyond classical contact processes or reaction processes, as only needs the existence of a (local) dimension reduction or mean-field technique in combination with a suitable two-parameter bifurcation unfolding.
}


\subsubsection*{Synchronization in phase oscillator networks}

\hl{
The Kuramoto model has been instrumental to understand the emergence of synchrony in coupled oscillator networks, ranging from synchronization of flashing fireflies to emergent neural synchrony~\cite{Strogatz2000,Acebron2005}. 
For a network of~$N$ Kuramoto oscillators, the state of oscillator $k\in\{1,\dotsc,N\}$ is given by the phase variable $\theta_k\in\R/(2\pi\Z)$. The phases evolve according to
\begin{equation}\label{eq:Kuram}
\dot\theta_k = \omega_k + \frac{K_2}{N}\sum_{j=1}^N\sin(\theta_j-\theta_k),
\end{equation}
for $k=1,\dotsc,N$ where~$K_2$ is the coupling strength between oscillators and the intrinsic frequencies~$\omega_k$ sampled from a unimodal distribution. 
Write $\txti:=\sqrt{-1}$. The (complex-valued) Kuramoto order parameter $Z = R\txte^{\txti\phi} = \frac{1}{N}\sum_{j=1}^N \txte^{\txti\theta_j}$ describes the mean field of the oscillators. Specifically, its absolute value encodes the level of synchrony in the system: $R=0$ if the phases all oscillators are evenly distributed around the circle implies, and $R=1$ if all oscillators are phase synchronized.
This classical model exhibits---in analogy to phase transitions in statistical mechanics---a second-order transition from an incoherent state to a partially coherent state as the coupling strength~$K_2$ between oscillators is increased~\cite{Daido1990}.
}

Kuramoto oscillators interact pairwise, so a natural generalization is to consider the additional effect of non-pairwise interactions~\cite{Rosenblum2007,Tanaka2011a,Bick2017c} since they arise naturally in phase reductions of coupled nonlinear oscillators~\cite{Ashwin2015a,Leon2019a}.
Skardal and Arenas~\cite{Skardal2019b} showed that first-order transitions to synchrony arise in a variation of the Kuramoto model with non-additive triplet interactions where the phase of oscillator~$k$ evolves according to
\begin{equation}
\begin{split}\label{eq:KuramNP}
\dot\theta_k &= \omega_k + \frac{K_2}{N}\sum_{j=1}^N\sin(\theta_j-\theta_k)
+\frac{K_3}{N}\sum_{j,l=1}^N\sin(2\theta_l-\theta_j-\theta_k),
\end{split}
\end{equation}
and the intrinsic frequencies~$\omega_k$ are sampled from a Lorentzian distribution with mean~$0$ and width~$1$; the choice of parameter for the Lorentzian can be made without loss of generality by scaling time appropriately and going in a suitable co-rotating reference frame. The parameter~$K_2$ determines the strength of the additive interactions and~$K_3$ the strength of the triplet interactions. \hl{While~$K_3=0$ yields the classical Kuramoto model~\eqref{eq:Kuram} with a continuous synchronization transition, for sufficiently large~$K_3$ this transition can become discontinuous.}

\hl{The change to a discontinuous synchronization transition in phase oscillators with higher-order interactions can be understood in terms of the universal route described above.}
Let $\bar w$ denote the complex conjugate of a complex number~$w$.
In the mean-field limit of $N\to\infty$ oscillators, the network dynamics~\eqref{eq:KuramNP} can be described using the Ott--Antonsen reduction~\cite{Ott2008,Bick2018c}: In the limit, the network dynamics are (exactly) described by the ODE
\begin{align*}
\dot Z & = -Z+\frac{1}{2}\left((K_2Z+K_3Z^2\bar{Z}) - (K_2\bar{Z}+K_3\bar{Z}^2{Z})Z^2\right),
\end{align*}
of the order parameter~$Z$; the derivation is given explicitly in~\cite{Skardal2019b}. Substituting polar coordinates $Z = R\txte^{\txti\phi}$ the dynamics of mean-field phase~$\phi$ and mean-field amplitude~$R$ decouple, yielding the effectively one-dimensional dynamics
\begin{align*}
\dot R &= \left(\frac{K_2}{2}-1\right)R + \left(\frac{K_3}{2}-\frac{K_2}{2}\right)R^3 -\frac{K_3}{2}R^5.
\end{align*}
The nature of the synchronization transition is determined by the bifurcation of the equilibrium $R=0$.
Setting $p=\frac{K_2}{2}-1$, $q = \frac{K_3}{2}$ and $x=R$ directly yields the normal form expansion~\eqref{eq:NFsym} of the pitchfork bifurcation. The bifurcation of $R=0$ happens at $p=0$ ($K_2=2$) for any~$K_3$. For the Kuramoto model $K_3=0$, the pitchfork bifurcation is always supercritical (second-order). However, for $K_3>2$ the synchronization transition becomes a subcritical first-order transition in line with our universal approach.

\hl{
Many other variations to the Kuramoto model that change the nature of the synchronization transition~\cite{Leyva2013} and are likely to provide further examples of the universal route. For example, Zhang \emph{et al.}~\cite{Zhang2014} considered a modified Kuramoto model where the coupling depends on the intrinsic frequency. This yields coupling that suppresses oscillator synchronization as a dynamical analog to the Achlioptas rule for percolation in random networks~\cite{Achlioptas2009}. The Kuramoto model with this additional feature exhibits a discontinuous synchronization transition and relates our theory with explosive synchronization.
Other variations of the Kuramoto model that affect the synchronization transition include varying the properties of the intrinsic frequencies~\cite{Pazo2005,OmelChenko2012a} or generalized coupling structures that encode higher-order effects~\cite{Skardal2019b,Millan2019a}.}


\subsubsection*{Discontinuous percolation transitions}
\newcommand{\Ps}{P^*}
\newcommand{\ph}{\hat{p}}

\hl{The flow of fluids through porous media is an example of a percolation problem~\cite{Sethna2006}. In bond percolation, each link of an infinite lattice of nodes is occupied with probability~$\ph$ and connected nodes form clusters of connected nodes. The percolation probability~$P^*(\ph)$ describes whether a given node is part of an infinite cluster or not. Varying~$\ph$, the system undergoes a critical transition in terms of~$P^*(\ph)$---the percolation transition---as large-scale clusters emerge. By a classical result by Fortuin and Kasteleyn~\cite{Fortuin1972}, these percolation problems can also be understood in terms of the Potts model~\cite{Wu1982} as a generalization of the Ising model. This correspondence allows to relate the percolation transition and phase transitions in the Potts model.}

\hl{Whether the percolation transition is continuous or discontinuous now depends on the system parameters. For the Potts model on a Bethe lattice, the percolation probability can be evaluated using recursive relations~\cite{Peruggi1983}. Consider the $q$-state Potts model on a Bethe-lattice with coordination number~3. For $q=2$ this gives the Ising model. Now suppose that the bonds are occupied independently with homogeneous density~$\ph\in[0, 1]$. 
Evaluating the percolation probabilities recursively as a hierarchy of finite latices with percolation probability~$P_n$,} one obtains the percolation probability for the (infinite) lattice as a fixed point of the iteration
\begin{equation}\label{eq:PercProb}
P_{n+1} = \frac{2\hat pP_n+(q-2)\ph^2P_n^2}{1+\ph^2(q-1)P_n^2}=:H(P_n)
\end{equation}
as calculated in~\cite{Chayes1986}.
The percolation transition of the fixed point~$\Ps=0$ of~$H$ happens at the critical bond density~$\hat p = \frac{1}{2}$; whether this transition in varying~$\ph$ is continuous or discontinuous depends on the number of states~$q$ of the \hl{Potts model~\cite{Chayes1986}: If $1\leq q\leq 2$ the transition is continuous and if $q>2$ the transition is discontinuous.}

This change of criticality of the percolation transition can be understood within the general dynamical framework introduced above. Set $p:=\hat p - \frac{1}{2}$ and consider the ODE
\begin{equation}\label{eq:PercProbCont}
\dot x = f(x,p,q) := 
H(x)-x
\end{equation}
obtained by seeing~\eqref{eq:PercProb} as a difference equation.
By definition, the fixed points of~$H$ in~\eqref{eq:PercProb} correspond to equilibria of~$f$ in~\eqref{eq:PercProbCont}. Moreover, since $\partial_xf = \partial_xH-1$ and $\partial_xH>0$ in a neighborhood of $(x,p) = 0$, linear stability of stationary states coincides as well. Thus, the behavior of the percolation transition of~\eqref{eq:PercProb} is completely determined by the bifurcations of the equilibrium $\xs=0$ of~\eqref{eq:PercProbCont} at $p=0$. A Taylor expansion of $f(x,p,q)$ yields
\[
f(x,p,q) = 2xp + \frac{1}{4}(g(p)q-2g(p))x^2 + \dotsb
\]
with $g(p) = 4p^2+4p+1$. Thus, the change of criticality of the percolation transition at $q=2$ corresponds to a change from a supercritical to a subcritical transcritical bifurcation in the universal route described above.

How the the percolation probability changes with the bond density is also directly related to discontinuous transitions in the expected maximal cluster size of a random graph. Specifically, random graphs with an underlying hierarchical self-similar structure allow to calculate the percolation probability through recursive relations~\cite{Boettcher2009} as in the Potts model discussed above. By calculating the corresponding generating functions~\cite{Boettcher2012}, one can observe a discontinuous transition in the expected size of the largest cluster. 
\hl{Finally, the critical transition in the $q$-state Potts model has also inspired a rule that links classical bond percolation and a type of Achlioptas process~\cite{Achlioptas2009} by varying an additional parameter~\cite{Araujo2011}.
}

\section*{Discussion}

Our argument shows that from the perspective of bifurcation theory, one can expect a change 
from a continuous transition to a discontinuous transition as additional effects are added 
to a classical model. Here, we gave three explicit physical examples to illustrate
this universal route. Our formalism not only shows that a transition to an abrupt change of system properties is not surprising but also has the same underlying dynamical mechanism. In particular, our framework links the emergence of abrupt critical transitions for percolation, synchronization, and epidemic spreading explicitly through a common dynamical framework.

Many other variations---beyond the ones already discussed in the context of the explicit examples---are possible that fit into our framework. First, our mechanism also yields a natural explanation for first-order synchronization transitions observed in coupled nonlinear oscillators~\cite{Calugaru2018}, since higher-order effects affect the dynamics beyond the weak coupling limit~\cite{Leon2019a}. Second, we anticipate our theory to be relevant in neural networks. For example, networks of quadratic-integrate-and-fire neurons can be described by low-dimensional equations using a reduction closely related to the Ott--Antonsen approach~\cite{Montbrio2015,Bick2018c}. These equations show transcritical bifurcation in a limiting case~\cite{Pazo2016} that could shed light on the emergence of discontinuous transitions between low- and high firing dynamics~\cite{Schmidt2018}.
Finally, we expect our theory to apply also in further physical systems, for example, laser dynamics~\cite{Scott1975}, \hl{flocking~\cite{Guisandez2017a}}, or chemical reaction networks, where the same type of mechanism is bound to be relevant.

\section*{Methods}

The results are based on theoretical considerations.

\section*{Acknowledgments}

The authors would like to acknowledge numerous helpful comments by S.~Boccaletti, T.~Gross, and the anonymous referees, which helped improve the exposition of our results as well as embed them better into the broader context.
CB and CK gratefully acknowledge the support of the Institute for Advanced Study at the Technical University of Munich through a Hans Fischer Fellowship awarded to CB that made this work possible. 
CK acknowledges support via a Lichtenberg Professorship as well as support via the TiPES project funded the European Union’s Horizon 2020 research and innovation programme under grant agreement No.~820970.
CB acknowledges support from the Engineering and Physical Sciences Research Council (EPSRC) through the grant EP/T013613/1.
All data needed to evaluate the conclusions in the paper are present in the paper.


\begin{thebibliography}{10}

\bibitem{Haken1977}
H.~Haken, {\it {Synergetics}\/} (Springer, Berlin, Heidelberg, 1977).

\bibitem{Boccaletti2016}
S.~Boccaletti, {\it et~al.\/}, {\it Phys Rep\/} {\bf 660}, 1 (2016).

\bibitem{DSouza2019}
R.~M. D'Souza, J.~G{\'{o}}mez-Garde{\~{n}}es, J.~Nagler, A.~Arenas, {\it Adv in
  Phys\/} {\bf 68}, 123 (2019).

\bibitem{Pazo2005}
D.~Paz{\'{o}}, {\it Phys Rev E\/} {\bf 72}, 046211 (2005).

\bibitem{OmelChenko2012a}
O.~E. Omel'chenko, M.~Wolfrum, {\it Phys Rev Lett\/} {\bf 109}, 164101 (2012).

\bibitem{Skardal2019b}
P.~S. Skardal, A.~Arenas, {\it Comm Phys\/} {\bf 3}, 218  (2020).

\bibitem{Millan2019a}
A.~P. Mill{\'{a}}n, J.~J. Torres, G.~Bianconi, {\it Phys Rev Lett\/} {\bf 124},
  218301 (2020).

\bibitem{Zhang2014}
X.~Zhang, Y.~Zou, S.~Boccaletti, Z.~Liu, {\it Sci Rep\/} {\bf 4}, 5200 (2015).

\bibitem{GrossDLimaBlasius}
T.~Gross, C.~D. D'Lima, B.~Blasius, {\it Phys Rev Lett\/} {\bf 96}, (208701)
  (2006).

\bibitem{Iacopinietal}
I.~Iacopini, G.~Petri, A.~Barrat, V.~Latora, {\it Nat Commun\/} {\bf 10}, 2485
  (2019).

\bibitem{KuehnMC}
C.~Kuehn, {\it Control of Self-Organizing Nonlinear Systems\/}, E.~Sch{\"o}ll,
  S.~Klapp, P.~H{\"o}vel, eds. (Springer, 2016), pp. 253--271.

\bibitem{Bick2018c}
C.~Bick, M.~Goodfellow, C.~R. Laing, E.~A. Martens, {\it J Math Neurosci\/}
  {\bf 10}, 9 (2020).

\bibitem{GH}
J.~Guckenheimer, P.~Holmes, {\it Nonlinear Oscillations, Dynamical Systems, and
  Bifurcations of Vector Fields\/} (Springer, New York, NY, 1983).

\bibitem{Kielhoefer}
H.~Kielhoefer, {\it Bifurcation Theory: An Introduction with Applications to
  PDEs\/} (Springer, 2004).

\bibitem{Vanderbauwhede1992a}
A.~Vanderbauwhede, G.~Iooss, {\it Dynamics Reported\/} (Springer, 1992), pp.
  125--163.

\bibitem{Faye2018}
G.~Faye, A.~Scheel, {\it Trans Am Math Soc\/} {\bf 370}, 5843 (2018).

\bibitem{Landau2013}
L.~D. Landau, E.~M. Lifshitz, {\it {Statistical Physics}\/} (Elsevier Science,
  2013).

\bibitem{Pastor-Satorrasetal}
R.~Pastor-Satorras, C.~Castellano, P.~V. Mieghem, A.~Vespignani, {\it Rev Mod
  Phys\/} {\bf 87}, 925 (2015).

\bibitem{Pastor-SatorrasVespignani}
R.~Pastor-Satorras, A.~Vespignani, {\it Phys Rev Lett\/} {\bf 86}, 3200 (2001).

\bibitem{KuehnCT2}
C.~Kuehn, {\it J Nonlinear Sci\/} {\bf 23}, 457 (2013).

\bibitem{Janssen2004}
H.-K. Janssen, M.~M{\"{u}}ller, O.~Stenull, {\it Phys Rev E\/} {\bf 70}, 026114
  (2004).

\bibitem{Grassberger1982}
P.~Grassberger, {\it Z Phys B Con Mat\/} {\bf 47}, 365 (1982).

\bibitem{Strogatz2000}
S.~H. Strogatz, {\it Physica D\/} {\bf 143}, 1 (2000).

\bibitem{Acebron2005}
J.~Acebr{\'{o}}n, L.~Bonilla, C.~{P{\'{e}}rez Vicente}, F.~Ritort, R.~Spigler,
  {\it Rev Mod Phys\/} {\bf 77}, 137 (2005).

\bibitem{Daido1990}
H.~Daido, {\it J Stat Phys\/} {\bf 60}, 753 (1990).

\bibitem{Rosenblum2007}
M.~Rosenblum, A.~Pikovsky, {\it Phys Rev Lett\/} {\bf 98}, 064101 (2007).

\bibitem{Tanaka2011a}
T.~Tanaka, T.~Aoyagi, {\it Phys Rev Lett\/} {\bf 106}, 224101 (2011).

\bibitem{Bick2017c}
C.~Bick, {\it Phys Rev E\/} {\bf 97}, 050201(R) (2018).

\bibitem{Ashwin2015a}
P.~Ashwin, A.~Rodrigues, {\it Physica D\/} {\bf 325}, 14 (2016).

\bibitem{Leon2019a}
I.~Le{\'{o}}n, D.~Paz{\'{o}}, {\it Phys Rev E\/} {\bf 100}, 012211 (2019).

\bibitem{Ott2008}
E.~Ott, T.~M. Antonsen, {\it Chaos\/} {\bf 18}, 037113 (2008).

\bibitem{Leyva2013}
I.~Leyva, {\it et~al.\/}, {\it Sci Rep\/} {\bf 3}, 1281 (2013).

\bibitem{Achlioptas2009}
D.~Achlioptas, R.~M. D'Souza, J.~Spencer, {\it Science\/} {\bf 323}, 1453
  (2009).

\bibitem{Sethna2006}
J.~P. Sethna, {\it {Statistical Mechanics: Entropy, Order Parameters, and
  Complexity}\/}, Oxford Master Series in Physics (Oxford University Press,
  2006).

\bibitem{Fortuin1972}
C.~M. Fortuin, P.~W. Kasteleyn, {\it Physica\/} {\bf 57}, 536 (1972).

\bibitem{Wu1982}
F.~Y. Wu, {\it Rev Mod Phys\/} {\bf 54}, 235 (1982).

\bibitem{Peruggi1983}
F.~Peruggi, F.~di~Liberto, G.~Monroy, {\it J Phys A-Math Gen\/} {\bf 16}, 811
  (1983).

\bibitem{Chayes1986}
J.~T. Chayes, L.~Chayes, J.~P. Sethna, D.~J. Thouless, {\it Commun Math Phys\/}
  {\bf 106}, 41 (1986).

\bibitem{Boettcher2009}
S.~Boettcher, J.~L. Cook, R.~M. Ziff, {\it Phys Rev E\/} {\bf 80}, 041115
  (2009).

\bibitem{Boettcher2012}
S.~Boettcher, V.~Singh, R.~M. Ziff, {\it Nat Comm\/} {\bf 3}, 787 (2012).

\bibitem{Araujo2011}
N.~A.~M. Ara{\'{u}}jo, J.~S. Andrade, R.~M. Ziff, H.~J. Herrmann, {\it Phys Rev
  Lett\/} {\bf 106}, 095703 (2011).

\bibitem{Calugaru2018}
D.~Călugăru, J.~F. Totz, E.~A. Martens, H.~Engel, {\it Sci Adv\/} {\bf 6},
  eabb2637 (2020).

\bibitem{Montbrio2015}
E.~Montbri{\'{o}}, D.~Paz{\'{o}}, A.~Roxin, {\it Phys Rev X\/} {\bf 5}, 021028
  (2015).

\bibitem{Pazo2016}
D.~Paz{\'{o}}, E.~Montbri{\'{o}}, {\it Phys Rev Lett\/} {\bf 116}, 238101
  (2016).

\bibitem{Schmidt2018}
H.~Schmidt, D.~Avitabile, E.~Montbri{\'{o}}, A.~Roxin, {\it PLOS Comp Bio\/}
  {\bf 14}, e1006430 (2018).

\bibitem{Scott1975}
J.~Scott, M.~Sargent, C.~Cantrell, {\it Opt Commun\/} {\bf 15}, 13 (1975).

\bibitem{Guisandez2017a}
L.~Guisandez, G.~Baglietto, A.~Rozenfeld, {\it arXiv:1711.11531\/} pp. 1--11
  (2017).

\end{thebibliography}
\end{document}